# Rashba coupling and Lifshitz transition in monolayer graphene

PARTHA GOSWAMI

*Deshbandhu College, University of Delhi, Kalkaji, New Delhi-110019,India*

E-mail: physicsgoswami@gmail.com;Tel:0091-129-243-9099.

**Abstract.** We take a wide-angle view of the problem of monolayer graphene (MLG) where spin-degeneracy lifting is assumed to be possible by wedging in the tunable Rashba spin-orbit coupling(RSOC) and the sub-lattice staggered potential. We next consider the AB-stacked bi-layer graphene (BLG) system (the A-carbon of the upper sheet lying on top of the B-carbon of the lower one) and assume that a perpendicular electric field is created by the external gates deposited on the BLG surface. This system exhibits the occurrence of trigonal warping due to a (skew) interlayer hopping leading to the well-known Lifshitz transition(LT)[Y. Lemonik, I.L. Aleiner, C. Toke, and V.I. Fal'ko; arXiv:1006.1399]. We do not observe the replication of the features associated with BLG-LT in MLG in the presence of RSOC.

## 1. Introduction

In two very exhaustive review articles Castro Neto et al. **[1]** have discussed many intriguing properties of graphene. These distinctive features**[1,2]** have initiated a remarkably intensive study of electronic properties of graphene in recent years. The investigations are largely motivated by the quasi-relativistic nature of the single-particle excitation spectrum close to the Dirac points; in the monolayer graphene (MLG), the charge carriers are mass-less Dirac particles of chiral nature near these points characterized by vanishing density of states (DOS) and the near-zero chemical potential ($\mu$). Sufficiently far away from these points DOS is non-vanishing and $\mu \neq 0$. On the other hand, before the advent of graphene **[3]** on the horizon, one of the central elements of semiconductor spintronics**[4,5]**,viz. spin-orbit coupling (SOC), was under-going rapid development. The curiosity regarding the role of SOC in graphene emerged in due course. Following the pioneering work of Kane and Mele**[6]** where a novel type of quantum spin-Hall effect was predicted, in fact, there had been a significant interest in examining the SOC effect in graphene which culminated into quite a few benchmark studies **[7,8,9,10]**.

The SOC in graphene can be intrinsic and extrinsic types. The former, which corresponds the inherent asymmetry of electron hopping between next nearest neighbors, is weak (15- 30 meV) as the carbon nuclei is light; the weak hyperfine coupling is due to the fact that carbon materials consist predominantly of the nuclear spin free $^{12}C$ isotope. This makes the materials potentially a good spin conductor with long spin coherence times **[11]**. The latter, however, resembles the Rashba model **[12, 13]**, breaks the mirror symmetry **[6]**, and could be induced by tunable, external electric field perpendicular to the graphene sheet, or by electrostatic interaction with the substrate[**14**]. The latter also has been experimentally enhanced in graphene samples on Ni with intercalated Au atoms **[15]**. As reported by Castro Neto and F. Guinea **[16],** the extrinsic SOC could be also induced by impurities. The answer to the question whether graphene could have substantial extrinsic spin-orbit coupling (or, Rashba spin-orbit coupling (RSOC)), unfortunately, to date is still not satisfactory, even though tremendous efforts have been made to explore and improve the possibilities **[7,8,9,10,11,14,15,16]**. Bypassing this vital issue, we start with a (hypothetical) monolayer graphene (MLG) system under a strong, controlled gate voltage where the Rashba coupling is substantially larger(the relative Rashba coupling strength ($t_{Rashba}/t$)$\approx$ 0.1 where 't' is the first neighbor hopping term) than the intrinsic SOC**[10,17,18].** This is able to remove the spin degeneracy close to the **K** and **K′** points; RSOC is manifested through the non-isotropic spin-splitting of the bands (see Figure 1).We also include a sub-lattice staggered potential in our investigation. We show that the latter plays a supportive role in lifting the spin degeneracy in a description where one may be away from the neutrality points **K** and **K′**; close to these points, where the carriers are mass-less Dirac fermions, we notice that this potential is not required to play such role (see Figure 2). It would perhaps be of interest to note that the hypothetical system considered here is not the result of out-of-the-box thinking. There are real systems, such as a single molybdenum disulphide ($MoS_2$) **[19]** tri-layer – a direct gapped Dirac system, characterized by a strong SO coupling **[20,21]**. The atoms in the single layers of $MoS_2$ are arranged hexagonally as in graphene. There is, however, a subtle difference which has its observational consequences in the optical properties, viz., while graphene has an inversion centre, $MoS_2$ lacks it. Most importantly, in great contrast to graphene, $MoS_2$ is useful in device applications due to its semiconductor-like band gap.

The bi-layer graphene(BLG) system presents an entirely different landscape. The carriers, for example, in the (Bernal AB-stacked) BLG are neither Dirac nor Schrodinger fermions. In the Bernal stacking the two layers in BLG, consisting of two coupled honeycomb lattices with basis atoms ($A_1$, $B_1$ ) and ($A_2$, $B_2$) in the bottom and the top layers, respectively, are arranged in ($A_2$, $B_1$) fashion. That is, the A-carbon of the upper sheet lies on top of the B-carbon of the lower one. A (skew) interlayer hopping leads to a concurrent velocity $v_3$in addition to the Fermi velocity $v_F$. Due to the former, the system shows a topological change (Trigonal warping(TW)) in the Fermi surface density of states(DOS): The DOS splits into four pockets comprising of the central part and three legs**[22]**. Such splitting is an indication of the Lifshitz transition (LT)**[23]**. It may be mentioned that Rakyta et al.**[24]** have shown that an MLG system with RSOC exhibits TW. We find that the replication of the features associated with LT in BLG are absent in MLG. In sections 2 and 3 we present the results reasonably clearly relegating some calculation details to the appendix. The communication ends with some concluding remarks in section 4.

## 2. Rashba spin-orbit coupling

We write the following general Hamiltonian (H) of the monolayer graphene (MLG)in the basis $(a_{\mathbf{k}\uparrow}\,,\ b_{\mathbf{k}\uparrow}\,,\ a_{\mathbf{k}\downarrow},\ b_{\mathbf{k}\downarrow})$ in momentum space involving the Rashba spin-orbit coupling (and the intrinsic spin-orbit coupling):

$$H = \sum_{\mathbf{k}} (a^{\dagger}_{\mathbf{k}\uparrow}\ \ b^{\dagger}_{\mathbf{k}\uparrow}\ \ a^{\dagger}_{\mathbf{k}\downarrow}\ \ b^{\dagger}_{\mathbf{k}\downarrow})\, \hbar(\mathbf{k}) \begin{pmatrix} a_{\mathbf{k}\uparrow} \\ b_{\mathbf{k}\uparrow} \\ a_{\mathbf{k}\downarrow} \\ b_{\mathbf{k}\downarrow} \end{pmatrix}, \quad (1)$$

$$\hbar(\mathbf{k}) = \begin{pmatrix} -t_{so}\gamma_{so} + M + V & -t\gamma_0 & 0 & t_R(\gamma_{R1} - \gamma_{R2}) \\ -t\gamma^*_0 & t_{so}\gamma_{so} + M - V & t_R(-\gamma^*_{R1} - \gamma^*_{R2}) & 0 \\ 0 & t_R(-\gamma_{R1} - \gamma_{R2}) & t_{so}\gamma_{so} - M + V & -t\gamma_0 \\ t_R(\gamma^*_{R1} - \gamma^*_{R2}) & 0 & -t\gamma^*_0 & -t_{so}\gamma_{so} - M - V \end{pmatrix}$$

where the Hamiltonian focuses on the π-orbitals only. In Eq.(2) M, and V, respectively, correspond to the exchange field term, and the staggered AB sub-lattice potential. Besides, $\gamma_0(\mathbf{k}) = [2\exp(ik_xa/2)\cos(\sqrt{3}k_ya/2) + \exp(-ik_xa)]$. The intrinsic spin-orbit coupling(ISOC) term, in coordinate representation, may be written as $H_{so} = (2it_{so}/\sqrt{3}) \sum_{ij} c^{\dagger}_{i\sigma}(\mathbf{s}.(\mathbf{d}_{kj}\times \mathbf{d}_{ik}))\, c_{j\sigma}$ where k is connecting the next-nearest neighbor sites i and j; $\mathbf{d}_{kj}$ is a unit lattice vector pointing from site j to site k. Here $c_{i\sigma}$ is π-orbital annihilation operator for an electron with spin σ on site i and **s** are the spin Pauli matrices. The Rashba spin-orbit coupling(RSOC) term, on the other hand, in coordinate representation may be written as $H_R = (it_R)\sum_{ij\mu\nu}[\,a^{\dagger}_{i\mu}(s_{\mu\nu}\times \mathbf{d}_{ij})_z\, b_{j\nu} - \text{h.c}]$ where once again **s** are the Pauli matrices representing the electron spin operator and μ, ν = 1, 2 denote the μν matrix elements of the Pauli matrices. The operators $a^{\dagger}_{i,\sigma}$ and $b^{\dagger}_{j,\sigma}$, respectively, correspond to the fermion creation operators with real spin σ for A and B sub-lattices in the mono-layer. Upon using the operator $A^{\dagger}_{\mathbf{k}} = (1/\sqrt{N})\sum_i \exp(i\mathbf{k}.\mathbf{R}_i)\, a^{\dagger}_{i,\sigma}$ where $\mathbf{R}_i$ is the Bravais vector of the ith unit cell and **k** lies in the first Brillouin zone (and similarly, introducing $B^{\dagger}_{\mathbf{k}}$ acting on sub-lattice B) it is easy to find that $H_R = (it_R)\sum_{\mathbf{k},\mu\nu}[A^{\dagger}_{\mathbf{k},\mu}(s_{\mu\nu}\times \mathbf{d}(\mathbf{k}))_z\, B_{\mathbf{k},\nu} - \text{h.c}]$ where $\mathbf{d}(\mathbf{k}) = -\sum_{j=1,2,3}\mathbf{d}_j\exp(-i\mathbf{k}.\boldsymbol{\delta}_j)$.For the nearest-neighbor hopping term(t), we have in the same representation $H_0 = -t\sum_{\mathbf{k},\sigma}\gamma_0(\mathbf{k})\, A^{\dagger}_{\mathbf{k},\sigma} B_{\mathbf{k},\sigma}$ with $\gamma_0(\mathbf{k}) = \sum_{j=1,2,3}\exp(-i\mathbf{k}.\boldsymbol{\delta}_j)$. The three nearest neighbor vectors are assumed to be $\boldsymbol{\delta}_1 = (a/2)(1,\sqrt{3})$ , $\boldsymbol{\delta}_2 = (a/2)(1,-\sqrt{3})$ ,and $\boldsymbol{\delta}_3 = a(-1,0)$ ; 'a' is the lattice constant. The second neighbor positions are given by $\mathbf{d}_{1,2} = \pm\mathbf{a}_1$, $\mathbf{d}_{3,4} = \pm\mathbf{a}_2$, $\mathbf{d}_{5,6} = \pm(\mathbf{a}_2 - \mathbf{a}_1)$ where $\mathbf{a}_1 = (a/2)(\ 3,\sqrt{3})$ and $\mathbf{a}_2 = (a/2)(\ 3,-\sqrt{3})$. We consider the term $\mathbf{d}(\mathbf{k}) = -\sum_{j=1,2,3}\mathbf{d}_j\exp(-i\mathbf{k}.\boldsymbol{\delta}_j) = -[\mathbf{d}_1\exp(-i\mathbf{k}.\boldsymbol{\delta}_1) + \mathbf{d}_2\exp(-i\mathbf{k}.\boldsymbol{\delta}_2) + \mathbf{d}_3\exp(-i\mathbf{k}.\boldsymbol{\delta}_3)]$. Upon simplification this may be written as $\mathbf{d}(\mathbf{k}) = d_1(k)\,\mathbf{i} + d_2(k)\,\mathbf{j}$ where

$$d_1(k) = -[(1/2)\{\exp(-ik_x(a/2) - ik_y(\sqrt{3}a/2))\} + (1/2)\{\exp(-ik_x(a/2) + ik_y(\sqrt{3}a/2))\} - \exp(ik_xa)], \quad (2)$$

$$d_2(k) = -[(\sqrt{3}/2)\{\exp(-ik_x(a/2) - ik_y(\sqrt{3}a/2))\} - (\sqrt{3}/2)\{\exp(-ik_x(a/2) + ik_y(\sqrt{3}a/2))\}]. \quad (3)$$

We now consider the term

$$(s_{\mu\nu}\times \mathbf{d}(\mathbf{k})) = (\sigma_x\,\mathbf{i} + \sigma_y\,\mathbf{j}) \times (d_1(k)\,\mathbf{i} + d_2(k)\,\mathbf{j})$$
$$= (\sigma_x\, d_2(k) - \sigma_y\, d_1(k))\,\mathbf{k}.$$

With the aid of the right-hand-side of this result we may write the following expression for $H_R$:

$$H_R = (it_R)\sum_{\mathbf{k},\mu\nu}[A^{\dagger}_{\mathbf{k},\mu}(s_{\mu\nu}\times \mathbf{d}(\mathbf{k}))_z\, B_{\mathbf{k},\nu} - \text{h.c}]$$
$$= (it_R)\sum_{\mathbf{k}}[(A^{\dagger}_{\mathbf{k}\uparrow}\ A^{\dagger}_{\mathbf{k}\downarrow})\begin{pmatrix} 0 & d_2 + id_1 \\ d_2 - id_1 & 0 \end{pmatrix}\begin{pmatrix} B_{\mathbf{k}\uparrow} \\ B_{\mathbf{k}\downarrow} \end{pmatrix}$$
$$- (B^{\dagger}_{\mathbf{k}\uparrow}\ B^{\dagger}_{\mathbf{k}\downarrow})\begin{pmatrix} 0 & d^*_2 + id^*_1 \\ d^*_2 - id^*_1 & 0 \end{pmatrix}\begin{pmatrix} A_{\mathbf{k}\uparrow} \\ A_{\mathbf{k}\downarrow} \end{pmatrix}]$$
$$= t_R\sum_{\mathbf{k}}[(i\,d_2 - d_1)\, A^{\dagger}_{\mathbf{k}\uparrow} B_{\mathbf{k}\downarrow} + (i\,d_2 + d_1)\, A^{\dagger}_{\mathbf{k}\downarrow} B_{\mathbf{k}\uparrow}$$
$$+ (-i\,d^*_2 + d^*_1)\, B^{\dagger}_{\mathbf{k}\uparrow} A_{\mathbf{k}\downarrow} + (-i\,d^*_2 - d^*_1)\, B^{\dagger}_{\mathbf{k}\downarrow} A_{\mathbf{k}\uparrow}]. \quad (4)$$

We obtain

$$(i\,d_2 - d_1) = \gamma_{R,1} - \gamma_{R,2}\,,\ (i\,d_2 + d_1) = -\gamma_{R,1} - \gamma_{R,2},$$

$$(-i\,d^*_2 + d^*_1) = -\gamma^*_{R,1} - \gamma^*_{R,2}\,,\ (-i\,d^*_2 - d^*_1) = \gamma^*_{R,1} - \gamma^*_{R,2}.$$

where

$$\gamma_{R,1} = [\exp(-i\,k_xa/2)\cos(\sqrt{3}k_ya/2) - \exp(i\,k_xa)],$$

$$\gamma_{R,2} = \sqrt{3}\exp(-i\,k_xa/2)\sin(\sqrt{3}k_ya/2). \quad (5)$$

In the absence of intrinsic spin-orbit coupling (ISOC) and the exchange field, the eigenvalues (ε) of the Hamiltonian in (1) are given by a bi-quadratic equation in ε:

$$\varepsilon^4 - 2\varepsilon^2\{t^2|\gamma_0|^2 + \tfrac{1}{2}t_R^2(|\gamma_{R,1} + \gamma_{R,2}|^2) + \tfrac{1}{2}t_R^2(|\gamma_{R,1} - \gamma_{R,2}|^2) + V^2\} - f(k_x,k_y) = 0\,, \quad (6)$$

$$f(k_x,k_y) \equiv [\,t^2t_R^2\, g(k_x,k_y) - \{V^4 + 2t^2|\gamma_0|^2V^2 + V^2t_R^2(|\gamma_{R,1} + \gamma_{R,2}|^2) + V^2t_R^2(|\gamma_{R,1} - \gamma_{R,2}|^2) + t^4|\gamma_0|^4 + t_R^4(|\gamma_{R,1} + \gamma_{R,2}|^2)(|\gamma_{R,1} - \gamma_{R,2}|^2)\}], \quad (7)$$

$$g(k_x,k_y) \equiv [\{16\cos(k_xa/2)\cos^3(\sqrt{3}k_ya/2) + 4\cos(5k_xa/2)\cos(\sqrt{3}k_ya/2) + 8\cos(k_xa)\cos^2(\sqrt{3}k_ya/2) + 24\cos(2k_xa)\sin^2(\sqrt{3}k_ya/2)\cos^2(\sqrt{3}k_ya/2)$$

$$+6\cos(k_xa)\sin^2(\sqrt{3}k_ya/2)$$

$$+12\cos(k_xa/2)\sin^2(\sqrt{3}k_ya/2)\cos(\sqrt{3}k_ya/2)\}$$

$$-\{8\cos(2k_xa)\cos^4(\sqrt{3}k_ya/2)+2\cos(k_xa)$$

$$\cos^2(\sqrt{3}k_ya/2)+4\cos(k_xa/2)\cos^3(\sqrt{3}k_ya/2)$$

$$+8\cos(k_xa)\cos^2(\sqrt{3}k_ya/2)+2\cos(4k_xa)$$

$$+4\cos(5k_xa/2)\cos(\sqrt{3}k_ya/2)\}]. \quad (8)$$

It may be noted that if the ISOC and the exchange field terms are included there would be an additional term in (6) involving ε which makes the eigenvalue equation a quartic. A quartic may be solved for real ε's by Ferrari method given the suitable choice of the parameters. We note that the RSOC achieves the spin degeneracy lifting: It leads to four spin-split bands and the anisotropic band gap $G(k_x,k_y) = 2\,[f_0(k_x,k_y) - g_0(k_x,k_y)]^{1/2}$ between the two bands closer to $\varepsilon = 0$. Here $f_0(k_x,k_y) \equiv [t^2\,|\gamma_0|^2 + \frac{1}{2}\,t_R^2(|\gamma_{R,1}+\gamma_{R,2}|^2) + \frac{1}{2}\,t_R^2(|\gamma_{R,1}-\gamma_{R,2}|^2) + V^2]$ and $g_0(k_x, k_y) \equiv \sqrt{\{f_0(k_x,k_y)^2 + f(k_x,k_y)\}}$. With the suitable choice of the parameters (such as the relative Rashba coupling strength $(t_R/t) \approx 0.1$, and $(V/t) \approx 0.4$) the term under the radical sign in $G(k_x,k_y)$ will be positive. In the Figure 1(a) below, we have shown the four bands for these parameter values along $ak_y = \pm 2\pi/3\sqrt{3}$. Upon lowering the parameter $(t_R/t)$ to about 0.03, the spin degeneracy lifting almost disappears as in Figure 1(b). On the other hand, upon changing $(V/t)$ from the critical value 0.4 to a lower value, the term under the radical sign in $G(k_x,k_y)$ will be negative. The latter underlines the supporting role that the sub-lattice staggered potential plays in lifting the spin degeneracy together with RSOC in a description where one may be away from the neutrality points $\mathbf{K}(2\pi/3a,\, 2\pi/3\sqrt{3}a)$ and $\mathbf{K'}(2\pi/3a,\, -2\pi/3\sqrt{3}a)$. If the system is placed on a certain substrate where there is a difference in the potential seen by the two atoms in the unit cell of graphene **[3]**, the chiral symmetry breaking occurs. We assume the tunability of RSOC additionally. The bands in Figure 1 clearly show that the linearization approximation which is to made below is valid only when one is close to the Dirac points where the chemical potential is zero; sufficiently far from the Dirac point (when the chemical potential is non-zero or the density of states is non-vanishing), the carriers no more behave as Dirac electrons.

We now carry out the linearization of the terms in $(\gamma_0(\mathbf{k}), \gamma_{R,1}, \gamma_{R,2})$ around the Dirac points $\mathbf{K}(2\pi/3a,\, 2\pi/3\sqrt{3}a)$ and $\mathbf{K'}(2\pi/3a,\, -2\pi/3\sqrt{3}a)$, respectively, writing $(k_xa = 2\pi/3 + \delta k_xa,\ k_ya = 2\pi/3\sqrt{3} + \delta k_ya)$ and $(k_xa = 2\pi/3 + \delta k_xa,\ k_ya = -2\pi/3\sqrt{3} + \delta k_ya)$. Around $\mathbf{K}(2\pi/3a,\, 2\pi/3\sqrt{3}a)$, we write $-t\,\gamma_0(\mathbf{k})$ as $-\sqrt{3}\,\hbar v_F\,|\delta\mathbf{k}|\,e^{i5\pi/6}\,exp(i\,\theta_k)$ where the Fermi velocity $\hbar v_F \approx (\sqrt{3}a|t|/2)$, $\cos(\theta_k) = \delta k_x/|\delta\mathbf{k}|$, $\sin(\theta_k) = \delta k_y/|\delta\mathbf{k}|$, and $\theta_k = \arctan(\delta k_y/\delta k_x)$. Similarly, around $\mathbf{K'}(2\pi/3a,\, -2\pi/3\sqrt{3}a)$, we write $k_xa = 2\pi/3 + \delta k_xa$ and $k_ya = -2\pi/3\sqrt{3} + \delta k_ya$ which yields the matrix element $-t\,\gamma_0(\mathbf{k}) = -\sqrt{3}\,\hbar v_F\,|\delta\mathbf{k}|\,e^{i5\pi/6}\,exp(-i\,\theta_k)$. The exercise with $(\gamma_{R,1},\gamma_{R,2})$ around $\mathbf{K}$, on the other hand, leads to $\gamma_{R,1} = [(3/4)\,(1-i\sqrt{3}) + (3\sqrt{3}a/8)\{\delta k_x + i\,\delta k_y\} + i \times (3a/8)\,(\delta k_x + i\,\delta k_y)]$ and $\gamma_{R,2} = = [(3/4)\,(1-i\sqrt{3}) - (3\sqrt{3}a/8)(\delta k_x + i\,\delta k_y) - i \times (3a/8)\,(\delta k_x + i\,\delta k_y)]$. Thus, the matrix element $t_R\,(\gamma_{R,1}(\mathbf{k}) - \gamma_{R,2}(\mathbf{k})) = \sqrt{3}\,\hbar v_3\, e^{i\pi/6}\,(\delta k_x + i\,\delta k_y)$ and $t_R\,(-\gamma_{R,1}(\mathbf{k}) - \gamma_{R,2}(\mathbf{k})) = -2\sqrt{3}\,\hbar v_3\, a^{-1}e^{-i\pi/3}$ where we have introduced a velocity $\hbar v_3 = \sqrt{3}\,at_R/2$. The corresponding terms around $\mathbf{K'}$ are $t_R\,(\gamma_{R,1}(k) - \gamma_{R,2}(k)) = 2\sqrt{3}\,\hbar v_3\, e^{-i\pi/3}$ and $t_R(-\gamma_{R,1}(k) - \gamma_{R,2}(k)) = -\sqrt{3}\,\hbar v_3\, e^{i\pi/6}\,|\delta k|\,exp(-i\theta_k)$. The calculation details are given in the appendix A below. It may be pointed out that the Hamiltonian (1) including the process carried out has close resemblance with that for the bi-layer graphene (BLG) system.

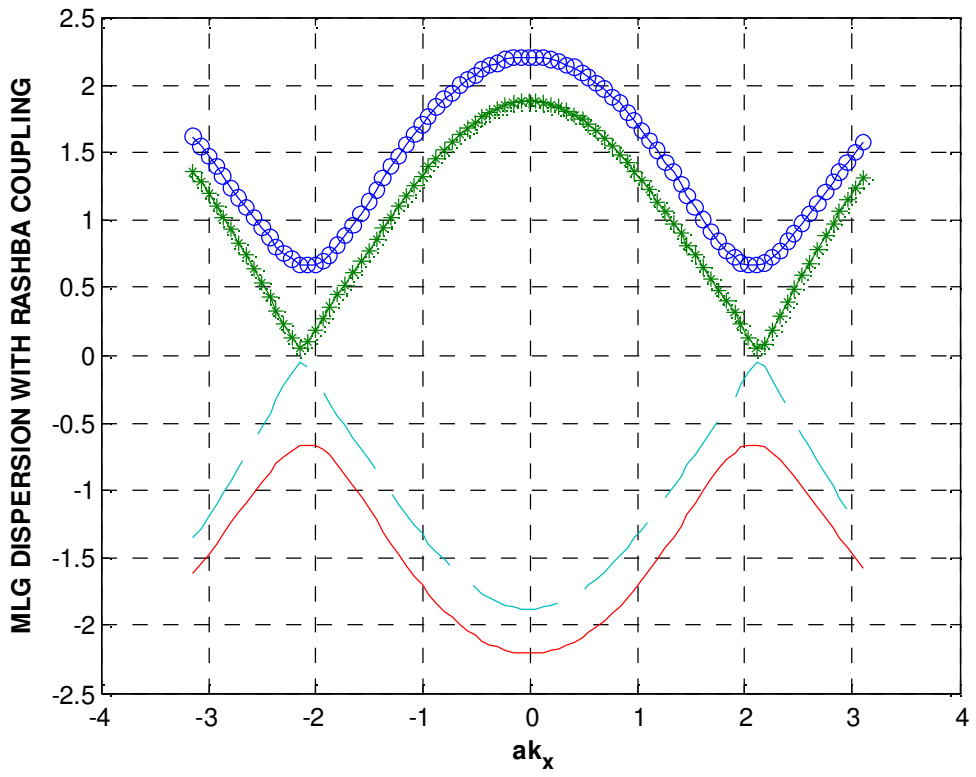

(a)

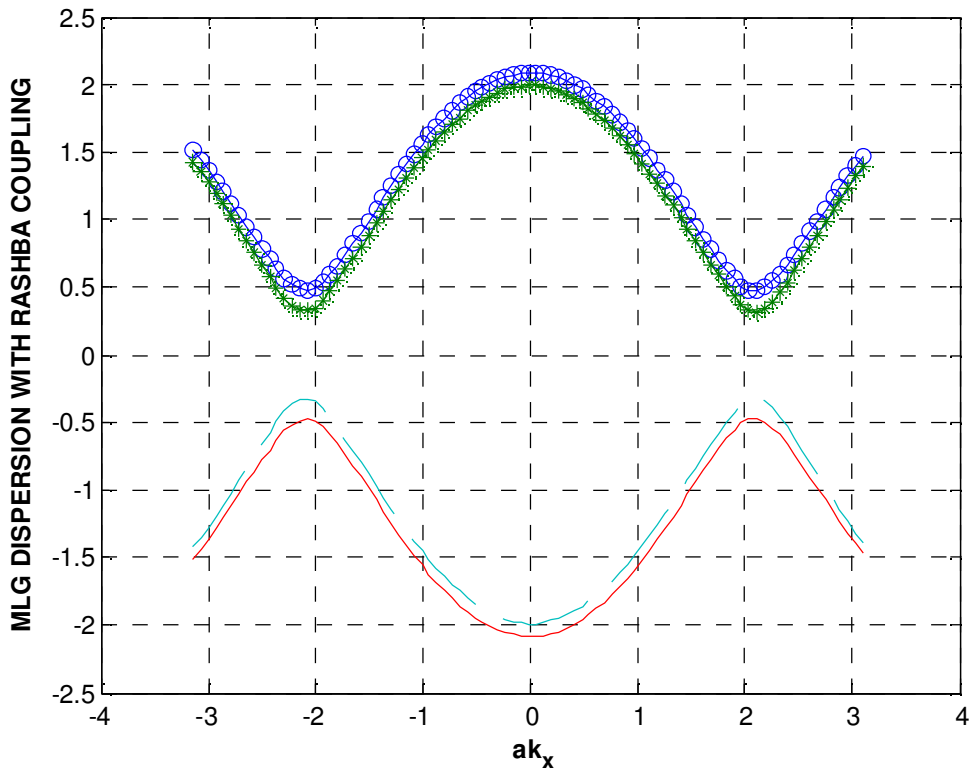

(b)

**Figure1.** The 2-D plot of the bands $\varepsilon(k_x,k_y)$ given by (6) as a function of $ak_x$ for $ak_y = \pm 2\pi/3\sqrt{3}$. (a) The parameter values are $(t_R/t) = 0.108$, and $(V/t) = 0.41$. (b) The parameter values are $(t_R/t) = 0.032$, and $(V/t) = 0.41$.

We are now in the position to write the following low-energy Dirac Hamiltonian matrices $\hbar_{\mathbf{K}}(\delta\mathbf{k})$ and $\hbar_{\mathbf{K'}}(\delta\mathbf{k})$ of the monolayer graphene (MLG)in the basis $(a_{\mathbf{k}\uparrow},\ b_{\mathbf{k}\uparrow},\ a_{\mathbf{k}\downarrow},\ b_{\mathbf{k}\downarrow})$ as in (1) and (2) in momentum space involving the Rashba spin-orbit coupling (and the sub-lattice staggered potential) around $\mathbf{K}$ and $\mathbf{K'}$. We have $\hbar_{\mathbf{K}}(\delta\mathbf{k}) =$

$$\begin{pmatrix} V & -\hbar v_F\,|\,\delta\mathbf{k}\,|\,e^{\frac{i5\pi}{6}}\,e^{i\theta_k} & 0 & \hbar v_3 e^{i\pi/6}\,|\,\delta\mathbf{k}\,|\,e^{i\theta_k} \\ -\hbar v_F\,|\,\delta\mathbf{k}\,|\,e^{\frac{-i5\pi}{6}}\,e^{-i\theta_k} & -V & -\,2a^{-1}\,\hbar v_3\,e^{i\pi/3} & 0 \\ 0 & -\,2a^{-1}\,\hbar v_3\,e^{-i\pi/3} & V & -\hbar v_F\,|\,\delta\mathbf{k}\,|\,e^{\frac{i5\pi}{6}}\,e^{i\theta_k} \\ \hbar v_3 e^{-i\pi/6}\,|\,\delta\mathbf{k}\,|\,e^{-i\theta_k} & 0 & -\hbar v_F\,|\,\delta\mathbf{k}\,|\,e^{\frac{-i5\pi}{6}}\,e^{-i\theta_k} & -V \end{pmatrix}$$

and $\hbar_{\mathbf{K}'}(\delta\mathbf{k})$ =

$$\begin{pmatrix} V & -\hbar v_F\,|\,\delta\mathbf{k}\,|\,e^{\frac{i5\pi}{6}}\,e^{-i\theta_k} & 0 & 2\,a^{-1}\hbar v_3\,e^{-i\pi/3} \\ -\hbar v_F\,|\,\delta\mathbf{k}\,|\,e^{\frac{-i5\pi}{6}}\,e^{i\theta_k} & -V & -\,\hbar v_3 e^{-i\pi/6}\,|\,\delta\mathbf{k}\,|\,e^{i\theta_k} & 0 \\ 0 & -\,\hbar v_3 e^{i\pi/6}\,|\,\delta\mathbf{k}\,|\,e^{-i\theta_k} & V & -\hbar v_F\,|\,\delta\mathbf{k}\,|\,e^{\frac{i5\pi}{6}}\,e^{-i\theta_k} \\ 2a^{-1}\,\hbar v_3\,e^{i\pi/3} & 0 & -\hbar v_F\,|\,\delta\mathbf{k}\,|\,e^{\frac{-i5\pi}{6}}\,e^{i\theta_k} & -V \end{pmatrix} \quad (6)$$

The eigenvalues(ε) of $\hbar_{\mathbf{K}}(\delta\mathbf{k})$ above are given by

$$\begin{aligned} \varepsilon^2 = {} & [(\hbar v_F\,|\,\delta\mathbf{k}\,|\,)^2 + V^2 + 2(a^{-1}\,\hbar v_3)^2 + (1/2)(\hbar v_3|\delta\mathbf{k}|\;)^2] \\ & \pm[(\tfrac{1}{4})(\hbar v_F|\delta\mathbf{k}|\;)^4 + 4(a^{-1}\,\hbar v_3\,)^4 + 4\,(\hbar v_3/a\,)^2(\hbar v_3|\delta\mathbf{k}|\;)^2 \\ & + (\hbar v_F|\delta\mathbf{k}|\;)^2\,(\hbar v_3|\delta\mathbf{k}|\;)^2 \\ & -4(\hbar v_F|\delta\mathbf{k}|\;)^2(a^{-1}(\,\hbar v_3\,)^2\,|\,\delta\mathbf{k}\,|\,)\cos(\theta_k - \pi/6) \\ & -2\,(\hbar v_3|\delta\mathbf{k}|\;)^2\,(\hbar v_3/a\,)^2\,]^{1/2}. \end{aligned} \quad (7)$$

The eigenvalues(ε) of $\hbar_{\mathbf{K}'}(\delta\mathbf{k})$ are given by

$$\begin{aligned} \varepsilon^2 = {} & [(\hbar v_F\,|\,\delta\mathbf{k}\,|\,)^2 + V^2 + 2(a^{-1}\,\hbar v_3)^2 + (1/2)(\hbar v_3|\delta\mathbf{k}|\;)^2] \\ & \pm[(\tfrac{1}{4})(\hbar v_F|\delta\mathbf{k}|\;)^4 + 4(a^{-1}\,\hbar v_3\,)^4 + 4\,(\hbar v_3/a\,)^2(\hbar v_3|\delta\mathbf{k}|\;)^2 \\ & + (\hbar v_F|\delta\mathbf{k}|\;)^2\,(\hbar v_3|\delta\mathbf{k}|\;)^2 \\ & -\,4(\hbar v_F|\delta\mathbf{k}|\;)^2\,(a^{-1}(\,\hbar v_3\,)^2\,|\,\delta\mathbf{k}\,|\,)\cos(\theta_k + \pi/6) \\ & -2\,(\hbar v_3|\delta\mathbf{k}|\;)^2\,(\hbar v_3/a\,)^2\,]^{1/2}. \end{aligned} \quad (8)$$

One may make use of the eigenvalues and the corresponding eigenvectors calculated (see Appendix B) to obtain the transport properties, such as the conductivity tensor components, using a Kubo formula **[25]**. We have obtained an exotic feature, viz. the Hall conductivity ($\sigma_H$) is non-zero even in the absence of a magnetic field, due to the fact that the two valleys **K** and **K′** are imbalanced in their contributions to $\sigma_H$ . We are, however, yet to ascertain whether the corresponding state is time reversal symmetry (TRS)-preserving quantum spin-Hall state (QSH) **[6]**, or TRS-breaking anomalous quantum Hall (AQH) state **[18]**.

Close to the Dirac points **K**($2\pi/3a$, $2\pi/3\sqrt{3}a$) and **K′** ($2\pi/3a$, $-2\pi/3\sqrt{3}a$) where the carriers are Dirac fermions, we find that the sub-lattice staggered potential is not required to play any more a supportive role in lifting the spin degeneracy. This is shown in the band plot in Figure 2 where the parameter values are $(t_R/t) = 0.108$, and $(V/t) = 0$. We also observe a remarkable fact with the bands close to ε = 0, viz. at a finite value of $(V/t) \sim 0.4$ starting with $(t_R/t) = 0$, where the 3-D band plots in the $\delta k_x$-$\delta k_y$ plane are symmetric about the lines $a\delta k_x = 0$ and $a\delta k_y = 0$, as we increase $(t_R/t)$ the plots exhibit distortion about $a\delta k_y = 0$ (see Figure 3). The distortion corresponding to **K** is in opposite sense to that corresponding to **K′.**

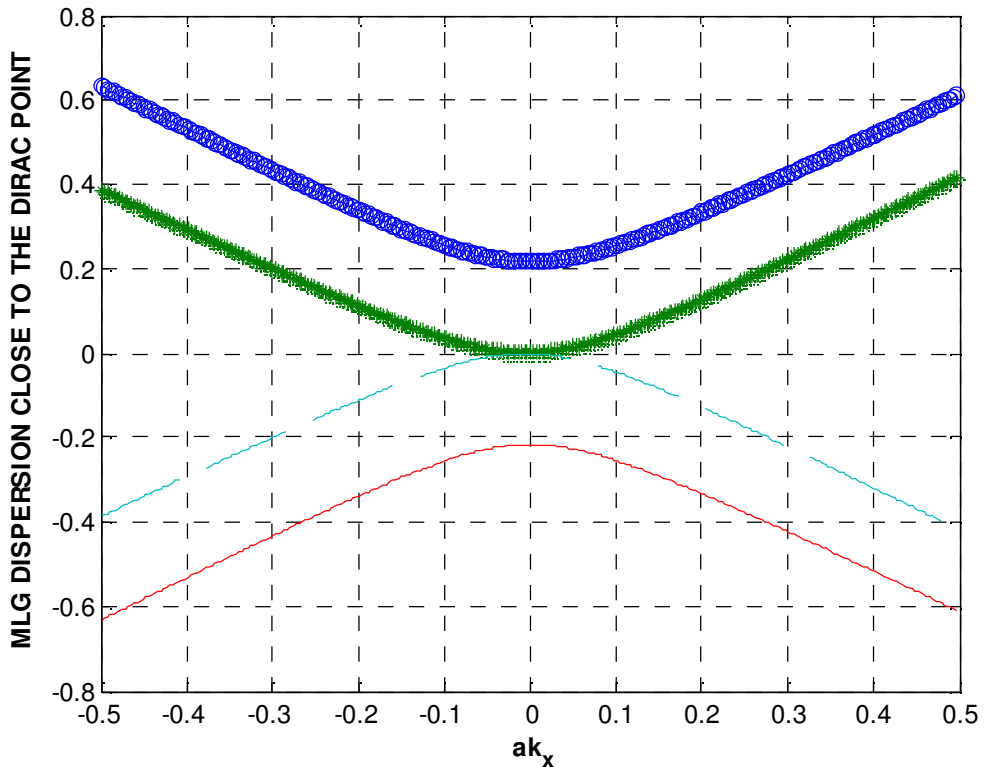

**Figure 2.** The MLG dispersion curves around Dirac point as a function of ($\delta k_x a$) along $\delta k_y a = 0$ for $(\hbar v_3/a\,)/\,(\hbar v_F/a\,) = 0.108$ and $(V/(\hbar v_F/a\,)=0$.

In the bi-layer graphene (BLG) system, with the inclusion of the skew interlayer hopping, one observes a topological change (Trigonal warping(TW)) in the Fermi surface density of states(DOS): The DOS splits into four pockets comprising of the central part and three legs**[22]**. Such splitting is an indication of the Lifshitz transition (LT)**[23]**. It may be mentioned that Rakyta et al.**[24]** have shown that an MLG system with RSOC exhibits TW. We discuss TW and LT in brief for the BLG system in the next section. As we have seen in Figure 3, these features are conspicuous by their absence in MLG.

### 3. Lifshitz transition in BLG

Close to the Dirac point in the Brillouin zone, the low-energy Hamiltonian **[1]** for the Bernal AB-stacked BLG

could be written in a compact form $H = \sum_{\delta k} \Psi^{\dagger}_{\delta k} H(\delta \mathbf{k}) \Psi_{\delta k}$ in the basis $(A_1, B_2, A_2, B_1)$ in the valley **K**. The row vector $\Psi^{\dagger}_{\delta k} = (a^{\dagger}_1(\delta \mathbf{k})\ b^{\dagger}_2(\delta \mathbf{k})\ a^{\dagger}_2(\delta \mathbf{k})\ b^{\dagger}_1(\delta \mathbf{k}))$ ;$a^{\dagger}_1(\delta \mathbf{k})$, $b^{\dagger}_2(\delta \mathbf{k})$, etc. stand for the fermion creation operators in the momentum space.

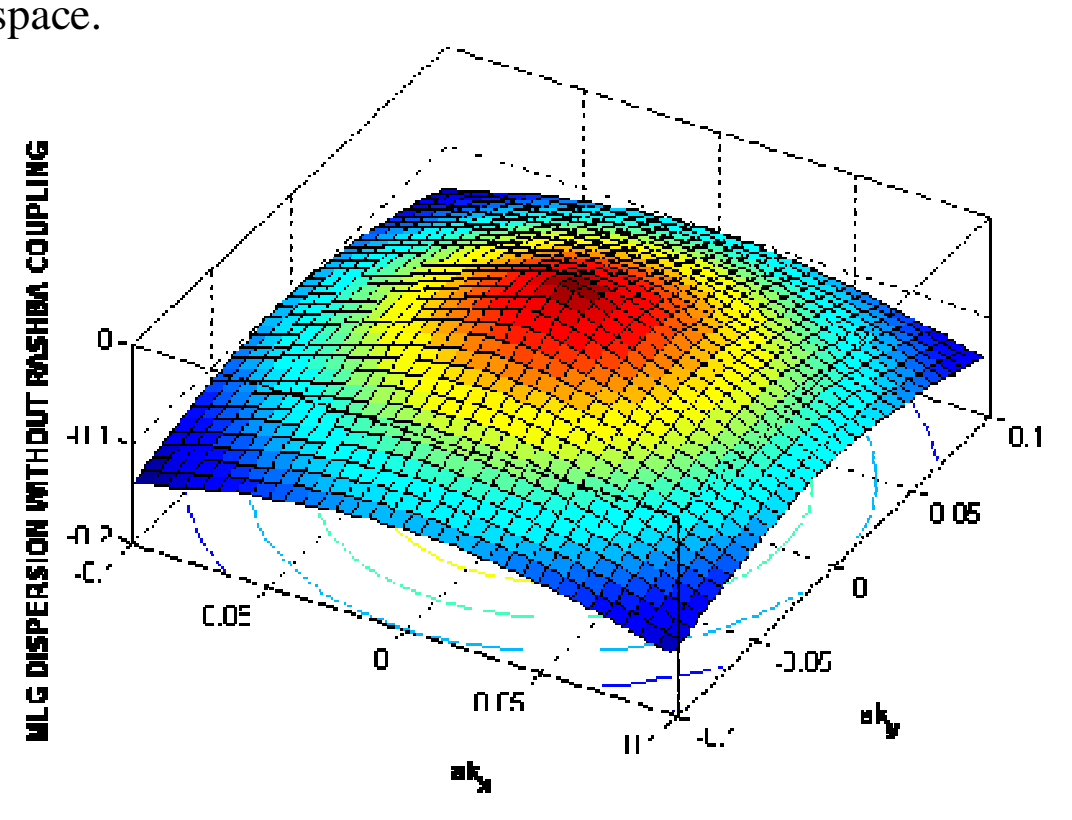

(a)

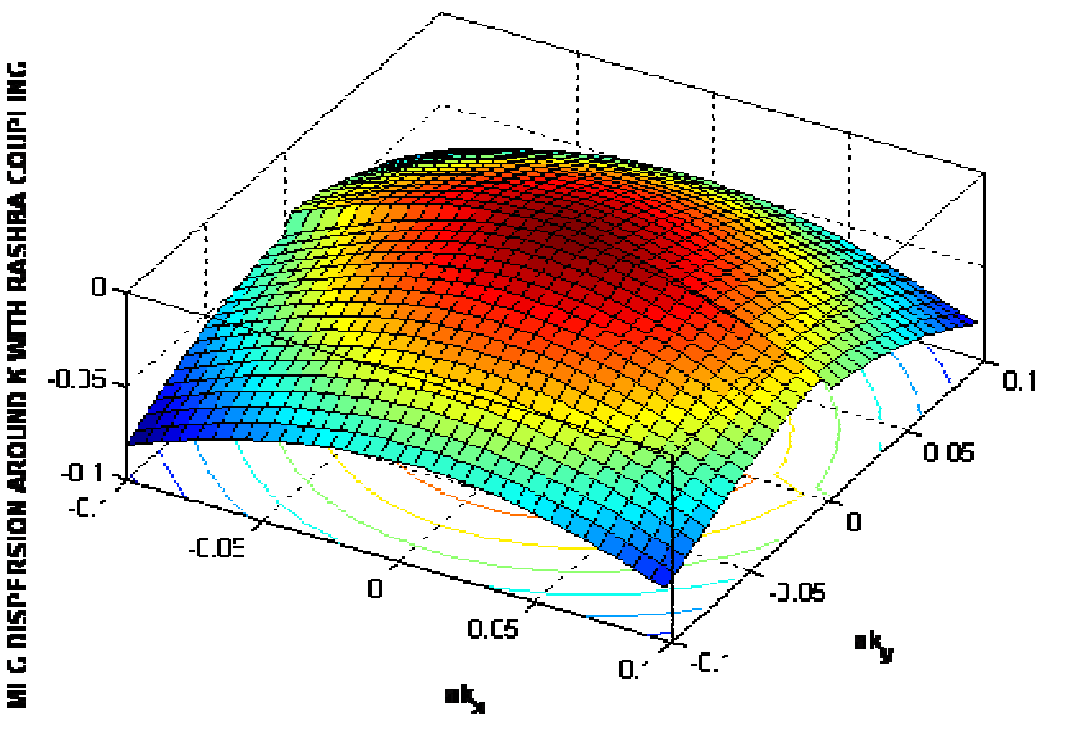

(b)

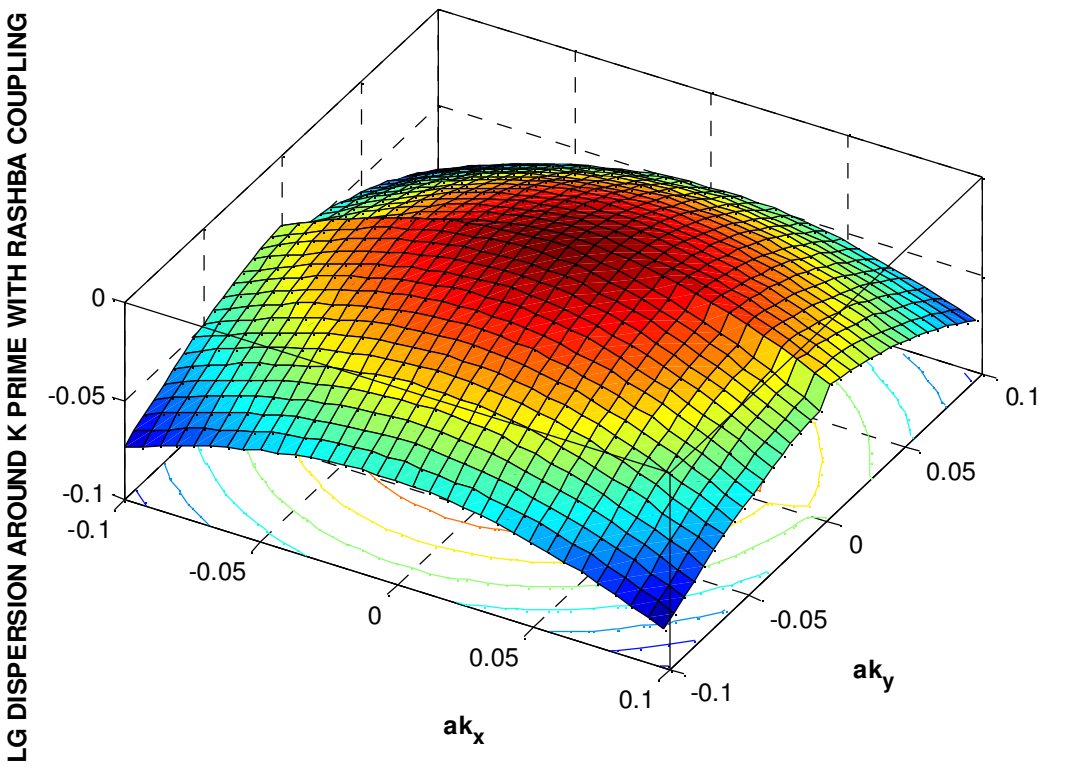

(c)

**Figure 3.** The MLG dispersion around Dirac points in the $\delta k_x$-$\delta k_y$ zone without/with RSOC. The parameter values are $(\hbar v_3/a)/(\hbar v_F/a) = 0.108$ and $(V/(\hbar v_F/a)) = 0$. (a) The valence band close to $\varepsilon = 0$ without RSOC. (b) The valence band close to $\varepsilon = 0$ around **K** with RSOC. (c) The valence band close to $\varepsilon = 0$ around **K′** with RSOC.

For the valley **K′,** the appropriate basis is $(B_2, A_1, B_1, A_2)$. We assume that a perpendicular electric field is (electrostatic bias V) created by the external gates deposited on the BLG surface. This induces a gap in the energy spectrum through a charge imbalance between the two graphene layers. The Hamiltonian matrix H ($\delta$**k**) is given by H ($\delta$**k**) **=** $\xi$ h($\delta$**k**) where h($\delta$**k**) is given in Appendix C. The meaning of the symbols in h($\delta$**k**) are as follows: $v_F$ is Fermi velocity (the speed of electrons in the vicinity of a Dirac point in the absence of interlayer hopping and is equal to $8\ 10^5 \text{m-s}^{-1}$), $\delta k = (\delta k_x + i\ \delta k_y)$ is a complex number and $\xi = \pm 1$; $\xi = +1$ corresponds to the valley **K** and $\xi = -1$ to the valley **K′**. Here $\gamma_1 = 0.39$ eV is the strongest interlayer coupling. There is an additional velocity $v_3 = 5.9\ \ 10^4\ \text{m-s}^{-1}$ due to a skew interlayer hopping. It causes a significant trigonal warping **[22]**of the energy dispersion.

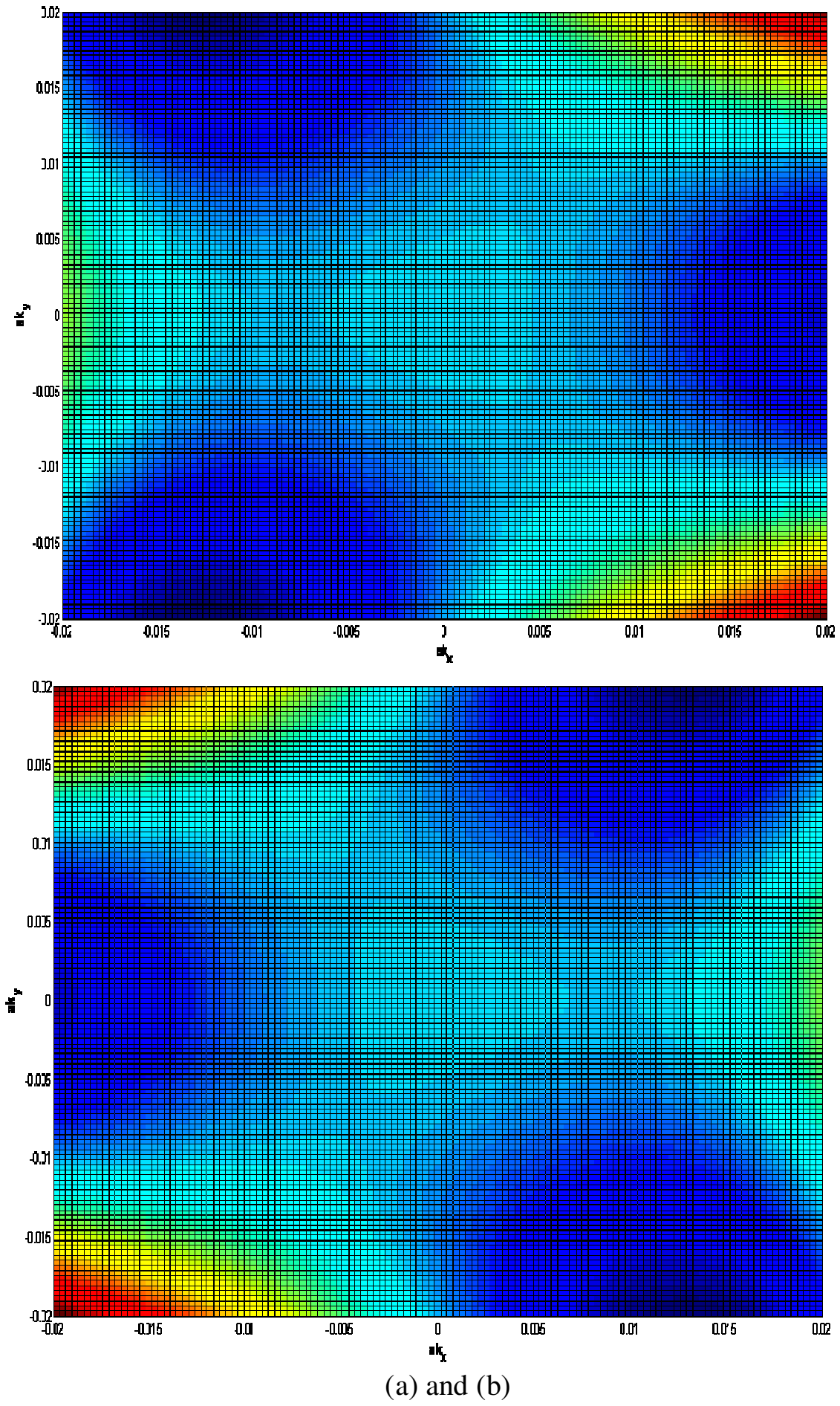

(a) and (b)

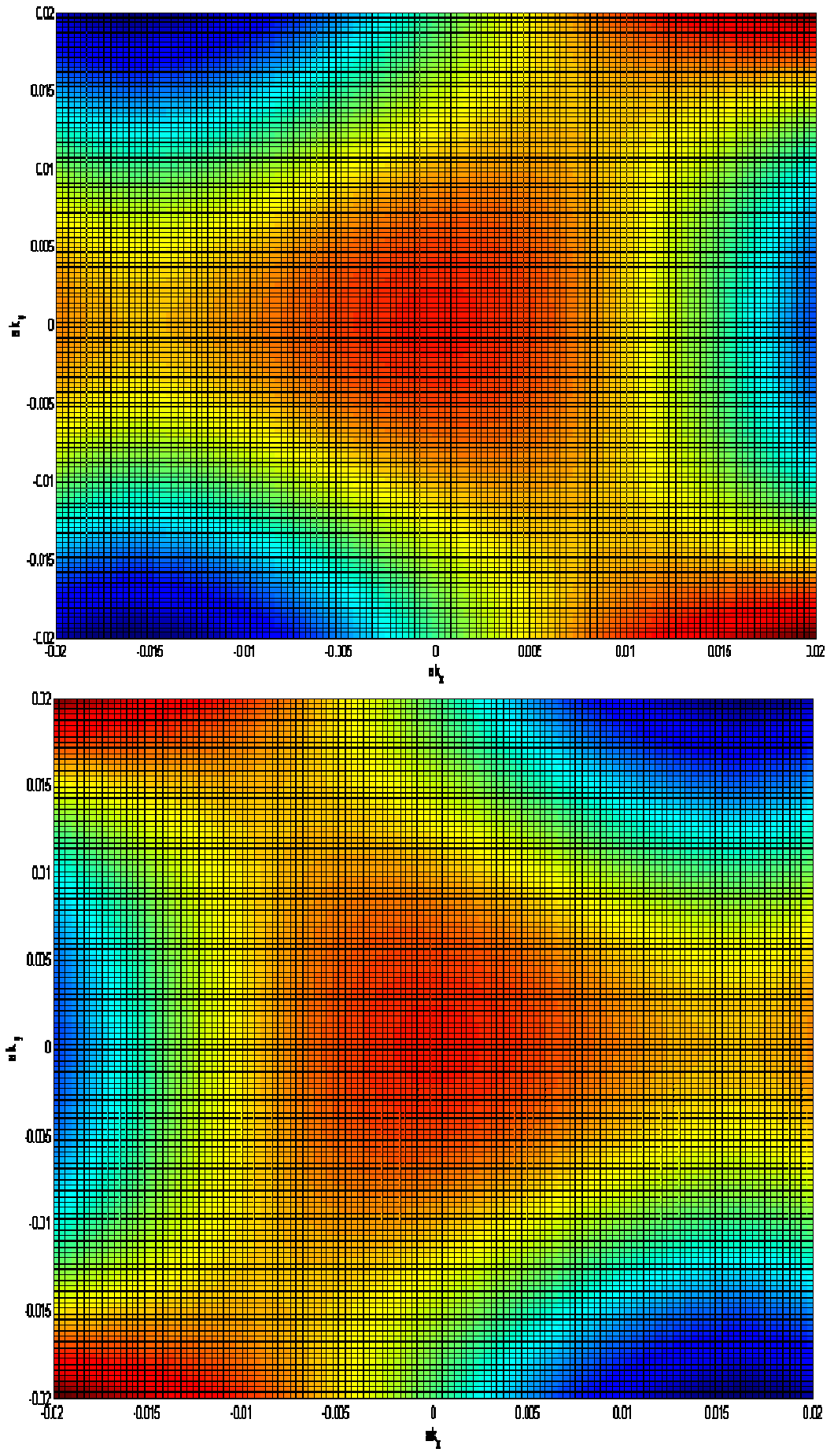

(c) and (d)

**Figure 4.** The contour plots of deformed lower energy band due to the inclusion of the last term in $\epsilon_2(\delta\mathbf{k})^4$. The figures ((a),(c)) and ((b),(d)) respectively, correspond to the valleys and $\mathbf{K'}$. The asymmetry of the plots at ((b),(d)) is inverted with respect to those at ((a),(c)). We see that the last term in $\epsilon_2(\delta\mathbf{k})^4$ correspond to the trigonal warping and split the band plot into four pockets comprising of the central part and three legs. For obtaining these contour plots, we have assumed $(v_F/a\gamma_1) = 7.9$. The remaining numerical values are $(V/\gamma_1) = 0.1$ [$(V/\gamma_1) = 0.14$]for (a) and(b) [(c)and(d)] and $(v_3/a\gamma_1) =$ 0. 7949. The scale of the plots is 0 to 1. Here $\gamma_1 = 0.39$ eV is the strongest interlayer coupling.

The energy eigen values $\varepsilon$ ($\delta\mathbf{k}$) of H ($\delta\mathbf{k}$) are given by a bi-quadratic equation. We obtain four bands, $\varepsilon_p^{\pm}$ ($\delta\mathbf{k}$), p = 1,2,with

$$\varepsilon_p(\delta\mathbf{k})^2 = [\epsilon_1(\delta\mathbf{k})^2 + V^2] +(-1)^p\sqrt{[\epsilon_1(\delta\mathbf{k})^4 + 4(v_F|\delta\mathbf{k}|)^2 V^2 - \epsilon_2(\delta\mathbf{k})^4]}, \quad (9)$$

where $\varepsilon_1$ and $\varepsilon_2$ , respectively, describes the lower and higher energy bands, and

$$\epsilon_1(\delta\mathbf{k})^2 = (v_F|\delta\mathbf{k}|)^2 + (1/2)(\gamma_1^2 + (v_3|\delta\mathbf{k}|)^2), \quad (10)$$

$$\epsilon_2(\delta\mathbf{k})^4 = (v_F|\delta\mathbf{k}|)^4 + (v_3|\delta\mathbf{k}|)^2\gamma_1^2 - 2 v_F^2\,\xi\,\gamma_1 v_3|\delta\mathbf{k}|^3\cos(3\varphi). \quad (11)$$

We have parameterized δk writing $\delta k_x = |\delta\mathbf{k}|\cos(\varphi)$ and $\delta k_y = |\delta\mathbf{k}|\sin(\varphi)$ which gives $\quad = |\delta\mathbf{k}|\exp(i\varphi)$.

The effect of the skew interlayer hopping, given by the last term in $\epsilon_2(\delta\mathbf{k})^4$ , on the four bands are found to be extremely sensitive to the bias. In Figure 4, where the band $\varepsilon_1$ ($\delta\mathbf{k}$) splits into four pockets comprising of the central part and three legs for $\varphi = \{0, 2\pi/3, 4\pi/3\}$, $\{\pi/3, \pi, 5\pi/3\}$, we have taken $(V/\gamma_1) = 0.1$ for (a) and (b) and $(V/\gamma_1) = 0.14$ for (c) and (d). In the former case the four pockets(cold patches) belong to lower energy while in the latter case the hot patches to slightly higher energy. We note that such splitting is an indication of the Lifshitz transition(a topological change in the Fermi surface) that is predicted in BLG. A higher value of $(V/\gamma_1)$, as much as 0.17 almost obliterates the four-pocket feature. Thus, the transition is bias-tunable. A comparison of the Figures 3 and 4 leads to the conclusion that the transitions in BLG and MLG(with RSOC) do not have the common features.

## 4. Concluding remarks

It is worth mentioning that, though the associated features of BLG-TW (Figure 4) and the distortion in the MLG bands (Figure 3) close to $\varepsilon = 0$ (due to the inclusion of RSOC) are found to be different, there is sufficient indication that MLG undergoes LT if RSOC strength ($t_R/t$) is quite strong(close to 0.1). The bands away from $\varepsilon = 0$, however, do not undergo significant deformation. The proper calculation of the impact of the distortion on the Fermi energy DOS requires the inclusion of the disorder and the many-body effects**[26]**. In a future investigation we hope to do so. Yet another important factor which has been overlooked here as follows: Since graphene is deposited or grown on some substrate, due to (the van der Waals) coupling with the latter which, in fact, corresponds to the absence of reflection symmetry in the plane of graphene sheet, long-wavelength strain (corrugations at low energies) would be generated on the sheet **[27]**. The Dirac spectrum survives under this reduction of symmetry only in the moderately strained case. The not-so-moderate strain, however, leads to opening of a gap in the excitation spectrum. A similar situation, i.e. manipulation of electronic properties by mechanical deformation, arises if suspended graphene is placed in external normal electric field. In particular, if the strain happens to be non-uniform, it acts as a pseudo-magnetic field[28,29,30]. Indeed, the vitality of the entire field of graphene physics rests upon such confounding issues.

## Appendix A

We carry out the linearization of the terms in (1) around **K** $(2\pi/3a, 2\pi/3\sqrt{3}a)$ and **K′** $(2\pi/3a, -2\pi/3\sqrt{3}a)$, respectively, writing ($k_xa = 2\pi/3 + \delta k_xa$, $k_ya = 2\pi/3\sqrt{3} + \delta k_ya$) and ($k_xa = 2\pi/3 + \delta k_xa$, $k_ya = -2\pi/3\sqrt{3} + \delta k_ya$). Around **K** $(2\pi/3a, 2\pi/3\sqrt{3}a)$, we have

$$\gamma_0(\mathbf{k}) = [2\exp(ik_xa/2)\cos(\sqrt{3}k_ya/2)+\exp(-ik_xa)]$$

$$= [2(\cos(\pi/3+\delta k_xa/2) + i\sin(\pi/3+\delta k_xa/2))\cos(\pi/3+\sqrt{3}\delta k_ya/2) +(\cos(2\pi/3+\delta k_xa) - i\sin(2\pi/3+\delta k_xa))]$$

$$= [\{2\times ½\times\cos(\delta k_xa/2) - 2\times\sqrt{3}/2\times\sin(\delta k_xa/2)\}$$
$$\times\{½\times\cos(\sqrt{3}\delta k_ya/2) - \sqrt{3}/2\times\sin(\sqrt{3}\delta k_ya/2)\}$$
$$+\{2i\times\sqrt{3}/2\times\cos(\delta k_xa/2) + 2i\times ½\times\sin(\delta k_xa/2)\}$$
$$\times\{½\times\cos(\sqrt{3}\delta k_ya/2) - \sqrt{3}/2\times\sin(\sqrt{3}\delta k_ya/2)\}$$
$$-½\times\cos(\delta k_xa) - \sqrt{3}/2\times\sin(\delta k_xa) - i\sqrt{3}/2\times\cos(\delta k_xa)$$
$$+ i\,½\times\sin(\delta k_xa)]$$

$$\approx ½\,[-(\sqrt{3}/2)\,\delta k_xa - (3/2)\delta k_ya - \sqrt{3}\,\delta k_xa$$
$$+ i\delta k_xa/2 - i\,(3\sqrt{3}/2)\delta k_ya + i\,\delta k_xa]$$

$$= ½\,[-(3\sqrt{3}/2)\,\delta k_xa + i3\delta k_xa/2 - (3/2)\delta k_ya - i\,(3\sqrt{3}/2)\delta k_ya]$$

$$= ½\,[-(3\sqrt{3}a/2)(\delta k_x+i\,\delta k_y) + i3a/2\,(\delta k_x+i\,\delta k_y)]$$

$$= (3ae^{i5\pi/6}/2)\,(\delta k_x+i\,\delta k_y). \qquad (A.1)$$

Thus, around **K,** the matrix element $-t\,\gamma_0(\mathbf{k}) = (-3ate^{i5\pi/6}/2)(\delta k_x+i\,\delta k_y) = -\sqrt{3}\,\hbar v_F\,e^{i5\pi/6}(\delta k_x+i\,\delta k_y)$.We introduce $\cos(\theta_k) = \delta k_x/|\delta\mathbf{k}|$, $\sin(\theta_k) = \delta k_y/|\delta\mathbf{k}|$, and $\theta_k = \arctan(\delta k_y/\delta k_x)$. This allows us to write $-t\,\gamma_0(\mathbf{k})$ as $-\sqrt{3}\,\hbar v_F\,|\delta\mathbf{k}|\,e^{i5\pi/6}(\cos(\theta_k) + i\sin(\theta_k)) = -\sqrt{3}\,\hbar v_F\,|\delta\mathbf{k}|\,e^{i5\pi/6}\,exp(i\,\theta_k)$ where the Fermi velocity $\hbar v_F \approx (\sqrt{3}a|t|/2)$. Similarly, around **K′** $(2\pi/3a, -2\pi/3\sqrt{3}a)$, we write ($k_xa = 2\pi/3 + \delta k_xa$ ; $k_ya = -2\pi/3\sqrt{3} + \delta k_ya$) and obtain the matrix element $-t\,\gamma_0(\mathbf{k}) = -\sqrt{3}\,\hbar v_F\,|\delta\mathbf{k}|\,e^{i5\pi/6}\,exp(-i\,\theta_k)$.

As for the matrix elements $\gamma_{R,1}(\mathbf{k})$ and $\gamma_{R,2}(\mathbf{k})$, around **K** $(2\pi/3a, 2\pi/3\sqrt{3}a)$, we write once again $k_xa = 2\pi/3 + \delta k_xa$ and $k_ya = 2\pi/3\sqrt{3} + \delta k_ya$ which gives

$$\gamma_{R,1}(\mathbf{k}) = [\exp(-i\,k_xa/2)\cos(\sqrt{3}k_ya/2) - \exp(i\,k_xa)]$$

$$=[(\cos(\pi/3+\delta k_xa/2) - i\sin(\pi/3+\delta k_xa/2))\cos(\pi/3+\sqrt{3}\delta k_ya/2)$$
$$-(\cos(2\pi/3+\delta k_xa) + i\sin(2\pi/3+\delta k_xa))]$$

$$= [\{½\times\cos(\delta k_xa/2) - \sqrt{3}/2\times\sin(\delta k_xa/2)\}$$
$$\times\{½\times\cos(\sqrt{3}\delta k_ya/2) - \sqrt{3}/2\times\sin(\sqrt{3}\delta k_ya/2)\}$$
$$-\{i\times\sqrt{3}/2\times\cos(\delta k_xa/2) + i\times ½\times\sin(\delta k_xa/2)\}$$
$$\times\{½\times\cos(\sqrt{3}\delta k_ya/2) - \sqrt{3}/2\times\sin(\sqrt{3}\delta k_ya/2)\}$$
$$+½\times\cos(\delta k_xa) + \sqrt{3}/2\times\sin(\delta k_xa) - i\sqrt{3}/2\times\cos(\delta k_xa)$$
$$+ i\,½\times\sin(\delta k_xa)]$$

$$\approx [(1/4) - (\sqrt{3}/8)\,\delta k_xa - (3/8)\delta k_ya - i\,\sqrt{3}/4 - i\times(1/8)\times\delta k_xa$$

$+ i\,(3\sqrt{3}/8)\delta k_y a + ½ + \sqrt{3}/2 \times \delta k_x a - i\sqrt{3}/2 + i\,½ \times (\delta k_x a)]$

$= [(3/4)\,(1-i\sqrt{3}) + (3\sqrt{3}a/8)\{\delta k_x + i\ \delta k_y\}$

$+ i \times (3a/8)\,(\delta k_x + i\,\delta k_y)]. \quad (A.2)$

$\gamma_{R,2}\,(\mathbf{k}) = \sqrt{3}\exp(-\,i\,k_x a/2)\sin(\sqrt{3}k_y a/2)$

$= \sqrt{3}\ \{\cos(k_x a/2) - i\sin(k_x a/2)\}\sin(\sqrt{3}k_y a/2)$

$= \sqrt{3}[\{\cos(\pi/3+\delta k_x a/2)\} - i\,\{\sin(\pi/3+\delta k_x a/2)\}]$

$\times \sin(\pi/3+\sqrt{3}\delta k_y a/2)$

$= \sqrt{3}[\{\ ½ \times \cos(\delta k_x a/2) - \sqrt{3}/2 \times \sin(\delta k_x a/2)\}$

$-\{i \times \sqrt{3}/2 \times \cos(\delta k_x a/2) + i \times ½ \times \sin(\delta k_x a/2)\}]$

$\times\{\sqrt{3}/2 \times \cos(\sqrt{3}\delta k_y a/2) + ½ \times \sin(\sqrt{3}\delta k_y a/2)\}$

$\approx \sqrt{3}[\ ½ - \sqrt{3}/4\,(\delta k_x a) - i \times \sqrt{3}/2 - i \times (1/4) \times (\delta k_x a)]$

$\times\{\sqrt{3}/2 + (\sqrt{3}/4)\,(\delta k_y a)\}$

$= \sqrt{3}\,[\sqrt{3}/4 - 3/8\,(\delta k_x a) - i \times 3/4 - i \times (\sqrt{3}/8) \times (\delta k_x a)$

$+ (\sqrt{3}/8) \times (\delta k_y a) - i \times 3/8(\delta k_y a)]$

$= [(3/4)\,(1-i\sqrt{3}) - (3\sqrt{3}a/8)(\delta k_x + i\ \delta k_y)$

$- i \times (3a/8)\,(\delta k_x + i\,\delta k_y)]. \quad (A.3)$

In view of (A.2) and (A.3) we may write

$\gamma_{R,1}(\mathbf{k}) - \gamma_{R,2}(\mathbf{k}) = (3\sqrt{3}a/4)(\delta k_x + i\,\delta k_y) + i \times (3a/4)\,(\delta k_x + i\,\delta k_y)$

$= (3a\,e^{i\pi/6}/2)(\delta k_x + i\,\delta k_y)\,, \quad (A.4)$

and

$-\gamma_{R,1}(\mathbf{k}) - \gamma_{R,2}\,(\mathbf{k}) = -\,(3/4)\,(1-i\sqrt{3}) - (3\sqrt{3}a/8)(\delta k_x + i\,\delta k_y)$

$- i \times (3a/8)\,(\delta k_x + i\,\delta k_y)$

$-\,(3/4)\,(1-i\sqrt{3}) + (3\sqrt{3}a/8)(\delta k_x + i\ \delta k_y)$

$+ i \times (3a/8)\,(\delta k_x + i\,\delta k_y)$

$= -\,3\,a\,a^{-1}\,e^{-i\pi/3}. \quad (A.5)$

Thus, around **K,** the matrix element $t_R\ (\gamma_{R,1}(\mathbf{k}) - \gamma_{R,2}(\mathbf{k})) = (3at_R e^{i\pi/6}/2)\ (\ \delta k_x + i\,\delta k_y\ ) = \sqrt{3}\ \hbar v_3\ e^{i\pi/6}\ (\ \delta k_x + i\,\delta k_y)$ where the skew velocity $\hbar v_3 = \sqrt{3}\ at_R/2$. We may now write $t_R\ (\gamma_{R,1}(\mathbf{k}) - \gamma_{R,2}(\mathbf{k})) = \sqrt{3}\ \hbar v_3\ e^{i\pi/6}\ |\delta\mathbf{k}|\ \ exp(i\,\theta_k)$. Similarly, $t_R\ (-\gamma_{R,1}(\mathbf{k}) - \gamma_{R,2}(\mathbf{k})) = -\,2\sqrt{3}\ \hbar v_3\ a^{-1}e^{-i\pi/3}$.

A similar calculation around $\mathbf{K'}$ $(2\pi/3a,\ -2\pi/3\sqrt{3}a)$ yields

$\gamma_{R,1}(\mathbf{k}) = [\exp(-i\,k_x a/2)\cos(\sqrt{3}k_y a/2) - \ \exp(i\,k_x a)]$

$= [\cos(\pi/3+\delta k_x a/2)\cos(-\pi/3+\sqrt{3}\delta k_y a/2)$

$- i\sin(\pi/3+\delta k_x a/2)\cos(-\pi/3+\sqrt{3}\delta k_y a/2)$

$-\cos(2\pi/3+\delta k_x a) - i\sin(2\pi/3+\delta k_x a))]$

$\approx\ [\{½ - (\sqrt{3}/4)\ \delta k_x a\} \times \{½ + (3/4)\delta k_y a\}$

$-\{i\,\sqrt{3}/2 + i \times ½ \times \delta k_x a/2\} \times \{½ + (3/4)\delta k_y a\}$

$+ ½ + \sqrt{3}/2 \times \delta k_x a - i\sqrt{3}/2 + i\,½ \times (\delta k_x a)]$

$=\ [\ (3/4)(1-i\,\sqrt{3}) + (3\sqrt{3}a/8)(\delta k_x - i\,\delta k_y)$

$+ i \times (3a/8)(\ \delta k_x - i\,\delta k_y)]$

$=\ [(3/2)\ e^{-i\pi/3} + (3a\ e^{i\pi/6}/4)(\delta k_x - i\,\delta k_y)]. \quad (A.6)$

$\gamma_{R,2}\,(\mathbf{k}) = \sqrt{3}\exp(-\,i\,k_x a/2)\sin(\sqrt{3}k_y a/2)$

$= \sqrt{3}[\{\cos(\pi/3+\delta k_x a/2)\} - i\,\{\sin(\pi/3+\delta k_x a/2)\}]$

$\times\ \sin(-\pi/3+\sqrt{3}\delta k_y a/2)$

$\approx \sqrt{3}[\ ½ - \sqrt{3}/4\,(\delta k_x a) - i \times \sqrt{3}/2 - i \times (1/4) \times (\delta k_x a)]$

$\times\{-\sqrt{3}/2 + (\sqrt{3}/4)\,(\delta k_y a)\}$

$= [(3/4)\,(-1+i\sqrt{3}) + (3\sqrt{3}a/8)(\delta k_x - i\ \delta k_y)$

$+ i \times (3a/8)\,(\delta k_x - i\,\delta k_y)]$

$= [(-3/2)\ e^{-i\pi/3} + (3a\ e^{i\pi/6}/4)\,(\delta k_x - i\ \ \delta k_y)]. \quad (A.7)$

In view of (A.6) and (A.7) we obtain

$\gamma_{R,1}(\mathbf{k}) - \gamma_{R,2}\,(\mathbf{k}) = 3\ a\ a^{-1}\ e^{-i\pi/3}$

and

$-\gamma_{R,1}(\mathbf{k}) - \gamma_{R,2}\,(\mathbf{k}) = -\,(3/2)\ e^{-i\pi/3} - (3a\ e^{i\pi/6}/4)(\delta k_x - i\,\delta k_y)$

$+ (3/2)\ e^{-i\pi/3} - (3a\ e^{i\pi/6}/4)\ (\delta k_x - i\ \ \delta k_y)$

$= -\,(3a\ e^{i\pi/6}/2)(\delta k_x - i\,\delta k_y). \quad (A.8)$

Thus, around **K′,** the matrix element $t_R\ (\gamma_{R,1}(\mathbf{k}) - \gamma_{R,2}(\mathbf{k})) = 2\sqrt{3}\ \hbar v_3\ e^{-i\pi/3}$ where the velocity $\hbar v_3 = \sqrt{3}\ at_R/2$. Also, $t_R\ (-\gamma_{R,1}(\mathbf{k}) - \gamma_{R,2}(\mathbf{k})) = -\ \sqrt{3}\ \hbar v_3\ e^{i\pi/6}\ |\ \delta k\ |\ exp(-i\,\theta_k)$.

## APPEDIX B

Around the valley **K**, the eigenvector, corresponding to an eigenvalue ‘ε’ in (7) is given by

$$
\begin{pmatrix}
e^{-i\theta_k/2} \\
[-2(\varepsilon - V)a^{-1}\hbar^2 v_3 v_F\, |\delta\boldsymbol{k}|\, e^{\frac{i\pi}{6}} e^{\frac{i\theta_k}{2}} + \hbar^2 v_3 v_F\, |\delta\boldsymbol{k}|\, |\delta\boldsymbol{k}|\, e^{\frac{i\pi}{3}} e^{-\frac{i\theta_k}{2}}]\, D \\
[(\varepsilon^2 - V^2)\hbar v_F |\delta\boldsymbol{k}| e^{\frac{-i\pi}{6}} e^{\frac{i\theta_k}{2}} - (\hbar v_F |\delta\boldsymbol{k}|)^3 e^{\frac{-i\pi}{6}} e^{\frac{i\theta_k}{2}} - 2\left(\frac{\hbar v_F}{a}\right)(\hbar v_3 |\delta\boldsymbol{k}|)^2 e^{-\frac{i\theta_k}{2}}]D \\
[(\varepsilon^2 - V^2)(\varepsilon - V)e^{-\frac{i\theta_k}{2}} - (\hbar v_F |\delta\boldsymbol{k}|)^2(\varepsilon - V)e^{-\frac{i\theta_k}{2}} - \left(\frac{2\hbar v_3}{a}\right)^2 (\varepsilon - V)e^{-\frac{i\theta_k}{2}}]D
\end{pmatrix}
$$

(B.1)

where

$$
D(\varepsilon) = \left[(\varepsilon^2 - V^2)\left(\hbar\, v_3 e^{i\theta_k}\, |\delta\boldsymbol{k}|\, e^{\frac{i\pi}{6}}\right) - 2(\hbar v_F |\delta\boldsymbol{k}|)^2\left(\frac{\hbar v_3}{a}\right)e^{2i\theta_k} - \left(\frac{2\hbar v_3}{a}\right)^2\left(\hbar\, v_3 e^{i\theta_k}\, |\delta\boldsymbol{k}|\, e^{\frac{i\pi}{6}}\right)\right].^{-1} \quad \text{(B.2)}
$$

As in (B.1) and (B.2), the eigenvector corresponding to an eigenvalue ‘ε’ in (8) around the valley $\mathbf{K'}$ is given by

$$
\begin{pmatrix}
e^{i\theta_k/2} \\
\left[2(\varepsilon - V)a^{-1}\hbar^2 v_3 v_F\, |\delta\boldsymbol{k}| e^{\frac{-i\pi}{6}} e^{\frac{i3\theta_k}{2}} - (\varepsilon - V)\hbar^2 v_3 v_F\, |\delta\boldsymbol{k}||\delta\boldsymbol{k}|\, e^{\frac{-i\pi}{3}} e^{\frac{i\theta_k}{2}}\right] D_1 \\
\left[(\varepsilon^2 - V^2)\hbar v_F |\delta\boldsymbol{k}| e^{\frac{-i\pi}{6}} e^{\frac{-i\theta_k}{2}} - (\hbar v_F |\delta\boldsymbol{k}|)^3 e^{\frac{-i\pi}{6}} e^{\frac{-i\theta_k}{2}} - 2\left(\frac{\hbar v_F}{a}\right)(\hbar v_3 |\delta\boldsymbol{k}|)^2 e^{\frac{i\theta_k}{2}}\right] D_1 \\
\left[(\varepsilon^2 - V^2)(\varepsilon - V)e^{\frac{i\theta_k}{2}} - (\hbar v_F |\delta\boldsymbol{k}|)^2(\varepsilon - V)e^{\frac{i\theta_k}{2}} - (\hbar v_3 |\delta\boldsymbol{k}|)^2\, (\varepsilon - V)e^{\frac{i\theta_k}{2}}\right] D_1
\end{pmatrix}
$$

(B.3)

where

$$
D_1(\varepsilon) = \left[\{(\varepsilon^2 - V^2) - (\hbar v_3 |\delta\boldsymbol{k}|)^2\}\left(2\hbar\, v_3 a^{-1}\, e^{\frac{-i\pi}{3}}\right) - (\hbar^2 v_3 v_F\, |\delta\boldsymbol{k}||\delta\boldsymbol{k}| e^{\frac{-i\pi}{3}})(\hbar v_F |\delta\boldsymbol{k}|)\quad e^{\frac{-i\pi}{6}} e^{-i\theta_k}\right]^{-1}.
$$

(B.4)

## APPENDIX C

The Hamiltonian matrix for the BLG system is given by h(δ**k**) =

$$
\begin{pmatrix}
V & v_3\delta\mathbf{k} & 0 & v_F\delta\mathbf{k}^* \\
v_3\delta\mathbf{k}^* & -V & v_F\,\delta\mathbf{k} & 0 \\
0 & v_F\,\delta\mathbf{k}^* & -V & \xi\gamma_1 \\
v_F\,\delta\mathbf{k} & 0 & \xi\gamma_1 & V
\end{pmatrix}.
$$

(C.1)

Here $v_F$ is Fermi velocity, and $\delta k = (\delta k_x + i\, \delta k_y)$ is a complex number. In the Bernal stacking the two layers in the bi-layer graphene, consisting of two coupled honeycomb lattices with basis atoms ($A_1$, $B_1$) and ($A_2$, $B_2$) in the bottom and the top layers, respectively, are arranged in ($A_2$, $B_1$) fashion. The intra-layer coupling between $A_1$ and $B_1$ and $A_2$ and $B_2$ is $\gamma_0$= 3.16 eV. The strongest interlayer coupling is between $A_2$ and $B_1$ with coupling constant $\gamma_1$= 0.39 eV. We consider a (skew) interlayer hopping between $A_1$ and $B_2$ with strength $\gamma_3$ = 0.315 eV. This introduces an additional velocity $v_3 = (3/2)a\,\gamma_3/\hbar = 5.9\times10^4$ m-s$^{-1}$. The values of these hopping integrals are taken to be same as in Ref. 1.